\documentclass[letterpaper]{ptephy}

\usepackage{boites}

\begin{document}

\title{Observation of the ``$K^-pp$"-like structure in the $d(\pi^+, K^+)$ reaction at 1.69~GeV/$c$ }

\author{\name{\fname{Yudai}~\surname{Ichikawa}}{1, 2}, \name{\fname{Tomofumi}~\surname{Nagae}}{1,*}, 
\name{\fname{Hiroyuki}~\surname{Fujioka}}{1}, 
\name{\fname{Hyoungchan}~\surname{Bhang}}{3}, \name{\fname{Stefania}~\surname{Bufalino}}{4}, \name{\fname{Hiroyuki}~\surname{Ekawa}}{1, 2}, \name{\fname{Petr}~\surname{Evtoukhovitch}}{5}, \name{\fname{Alessandro}~\surname{Feliciello}}{4},  \name{\fname{Shoichi}~\surname{Hasegawa}}{2}, \name{\fname{Shuhei}~\surname{Hayakawa}}{6}, \name{\fname{Ryotaro}~\surname{Honda}}{7}, \name{\fname{Kenji}~\surname{Hosomi}}{2}, \name{\fname{Kenichi}~\surname{Imai}}{2}, \name{\fname{Shigeru}~\surname{Ishimoto}}{8}, \name{\fname{Changwoo}~\surname{Joo}}{3}, \name{\fname{Shunsuke}~\surname{Kanatsuki}}{1}, \name{\fname{Ryuta}~\surname{Kiuchi}}{2}, \name{\fname{Takeshi}~\surname{Koike}}{7}, \name{\fname{Harphool}~\surname{Kumawat}}{9}, \name{\fname{Yuki}~\surname{Matsumoto}}{7}, \name{\fname{Koji}~\surname{Miwa}}{7}, \name{\fname{Manabu}~\surname{Moritsu}}{10}, \name{\fname{Megumi}~\surname{Naruki}}{1}, \name{\fname{Masayuki}~\surname{Niiyama}}{1}, \name{\fname{Yuki}~\surname{Nozawa}}{1}, \name{\fname{Ryosuke}~\surname{Ota}}{6}, \name{\fname{Atsushi}~\surname{Sakaguchi}}{6}, \name{\fname{Hiroyuki}~\surname{Sako}}{2}, \name{\fname{Valentin}~\surname{Samoilov}}{5}, \name{\fname{Susumu}~\surname{Sato}}{2}, \name{\fname{Kotaro}~\surname{Shirotori}}{10}, \name{\fname{Hitoshi}~\surname{Sugimura}}{2}, \name{\fname{Shoji}~\surname{Suzuki}}{8}, \name{\fname{Toshiyuki}~\surname{Takahashi}}{8}, \name{\fname{Tomonori}~\surname{Takahashi}}{10}, \name{\fname{Hirokazu}~\surname{Tamura}}{7}, \name{\fname{Toshiyuki}~\surname{Tanaka}}{6}, \name{\fname{Kiyoshi}~\surname{Tanida}}{3}, \name{\fname{Atsushi}~\surname{Tokiyasu}}{10}, \name{\fname{Zviadi}~\surname{Tsamalaidze}}{5}, \name{\fname{Bidyut}~\surname{Roy}}{9}, \name{\fname{Mifuyu}~\surname{Ukai}}{7}, \name{\fname{Takeshi}~\surname{Yamamoto}}{7} and \name{\fname{Seongbae}~\surname{Yang}}{3}}

\address{\affil{1}{Department of Physics, Kyoto University, Kyoto 606-8502, Japan}
\affil{2}{ASRC, Japan Atomic Energy Agency, Ibaraki 319-1195, Japan}
\affil{3}{Department of Physics and Astronomy, Seoul National University, Seoul 151-747, Korea}
\affil{4}{INFN, Istituto Nazionale di Fisica Nucleare, Sez. di Torino, I-10125 Torino, Italy}
\affil{5}{Joint Institute for Nuclear Research, Dubna, Moscow Region 141980, Russia}
\affil{6}{Department of Physics, Osaka University, Toyonaka 560-0043, Japan}
\affil{7}{Department of Physics, Tohoku University, Sendai 980-8578, Japan}
\affil{8}{High Energy Accelerator Research Organization (KEK), Tsukuba, 305-0801, Japan}
\affil{9}{Nuclear Physics Division, Bhabha Atomic Research Centre, Mumbai, India}
\affil{10}{Research Center for Nuclear Physics, Osaka University, Osaka 567-0047, Japan}
\email{nagae@scphys.kyoto-u.ac.jp}}

\begin{abstract}%
We have observed a ``$K^-pp$"-like structure in the $d(\pi^+,K^+)$
reaction at 1.69~GeV/$c$. In this reaction $\Lambda(1405)$ hyperon
resonance is expected to be produced as a doorway to form the $K^-pp$ through
the $\Lambda^*p\rightarrow K^-pp$ process.
However, most of the produced $\Lambda(1405)$'s would escape
from deuteron without secondary reactions.
Therefore, coincidence of high-momentum ($>$ 250~MeV/$c$) proton(s)
in large emission angles ($39^\circ<\theta_{lab.}<122^\circ$)
was requested to enhance the signal-to-background ratio.
A broad enhancement in the proton coincidence spectra are observed
around the missing-mass of 2.27~GeV/$c^2$, which corresponds
to the $K^-pp$ binding energy of 95~$^{+18}_{-17}$~(stat.)~$^{+30}_{-21}$~(syst.)~MeV and
the width of 162~$^{+87}_{-45}$~(stat.)~$^{+66}_{-78}$~(syst.)~MeV.
\end{abstract}

\subjectindex{Kaonic nuclei, J-PARC}

\maketitle
\begin{sloppypar}

\section{1. Introduction} 
Whether there exist kaonic nuclei or not is a key issue to make 
our understandings on the $\bar{K}N$ interaction in vacuum
and in nuclear medium. The information on the $\bar{K}N$
interaction has been obtained by analyzing low-energy
$\bar{K}N$ scattering data and kaonic-atom X-ray data~\cite{Martin81}.
The recent measurement of the energy shift and width on kaonic-hydrogen
X-ray in high precision by the SIDDHARTA group~\cite{SIDDHARTA} has
contributed a lot~\cite{Hyodo13}.
From the theoretical analyses by using these results, it is well known that
the $\bar{K}N$ interaction has a strong attraction in isospin 0 channel,
which suggests possible existence of kaonic bound state formations~\cite{AY02}.
Among them $K^-pp$ bound state composed of a $K^-$ and
two protons could be the simplest one, if existed.

Since this is a three-body system, several groups (see a recent summary in \cite{Gal13} and \cite{Maeda13, Bayar13, Revai13}) calculated
the binding energy and width of $K^-pp$ by applying various
few-body calculation techniques such as variational and
Faddeev type calculations.
The obtained binding energies are scattered in a broad range: 10\mbox{--}20~MeV for shallow potential cases
and 50\mbox{--}100~MeV for deep cases.
The width would be as wide as 70~MeV because of 
the strong $\bar{K}N$-$\pi\Sigma$ coupling. In addition there
could be non-mesonic absorption contributions of $\bar{K}NN\rightarrow\Lambda (\Sigma)N$.

The first experimental evidence of the $K^-pp$ bound state was reported
by the FINUDA collaboration~\cite{FINUDA05} in the stopped $K^-$
absorption reactions on $^6$Li, $^7$Li, and $^{12}$C targets.
They observed a lot of $\Lambda p$ pairs emitted in back-to-back,
and found the invariant mass of the pair significantly lower than $K^-pp$ mass threshold.
The binding energy of 115~$^{+6}_{-5}~$(stat.)~$^{+3}_{-4}$~(syst.)~MeV
and the decay width of $\Gamma =67~^{+14}_{-11}~$(stat.)$~^{+2}_{-3}$~(syst.)~MeV
were obtained.
However, there was a theoretical criticism~\cite{MORT} to interpret the observed
structure as the $K^-pp$ bound state.

Another experimental evidence was reported by the DISTO collaboration \cite{DISTO}.   
They measured the missng-mass and invariant mass spectra
in an exclusive reactions of $pp\rightarrow K^+\Lambda p$ at 2.85~GeV.
The binding energy of 103~$\pm$~3~(stat.)~$\pm$~5(syst.)~MeV
and the width of 118~$\pm$~8(stat.)~$\pm$~10(syst.)~MeV were obtained. 
However, they did not observe the signal at 2.50~GeV~\cite{DISTO2}, maybe due to the less 
production cross section of $\Lambda$(1405) at this energy.

There is also a report of a narrow and much deeper binding for the $K^-pp$ system observed in $\bar{p}$-$^4$He annihilations at rest~\cite{OBELIX}.
No peak was observed in the inclusive spectrum of the $\gamma d \rightarrow K^+\pi^-X$ reaction
at $E_{\gamma} = 1.5\mbox{--}2.4$~GeV \cite{TOKIYASU}.

Thus, the experimental situation  for the $K^-pp$ bound state is
not conclusive at this moment. It would be important to obtain new experimental
information in different reactions. 
In the J-PARC~E27 experiment, we used the $d(\pi^+, K^+)$ reaction at 1.69~GeV/$c$
to produce the $K^-pp$ system through the $\Lambda(1405)$ production
as a doorway~\cite{AYE27}. This is a simple production measurement with the smallest
final state effects.

\section{2. Experimental Setup}
The experiment was carried out at K1.8 beam line~\cite{Takahashi12} of the hadron
experimental hall at J-PARC~\cite{Nagamiya12}. 
In this beam line, separated $K^\pm, \pi^\pm, p,$ and $\bar{p}$
beams up to 2~GeV/$c$ are delivered. 
The detail of the experimental setup of this measurement are described in Ref.~\cite{YI_Jsympo, E27inclusive}.

The beam line is equipped with
a beam line spectrometer for the incident $\pi^+$ momentum reconstruction
composed of four quadrupole magnets and one dipole magnet.
The out-going $K^+$ momentum was reconstructed with the 
Superconducting Kaon Spectrometer (SKS) with the
momentum resolution of $\Delta p/p\sim 1\times 10^{-3}$.

A liquid hydrogen/deuterium target was installed 1.3~m upstream from the
entrance of the SKS magnet, so that the solid angle acceptance of the
SKS was about 100~msr. The size of the target cell was 120~mm in length
and 67.3~mm in diameter, which contains 1.99~g/cm$^2$
of liquid deuterium.

\section{2.1. Range Counter Arrays}
\begin{figure}
\vspace*{-0.7cm} 
 \begin{minipage}{0.5\hsize}
  \begin{center}
 \hspace*{-0.7cm}
     \includegraphics[width=65mm]{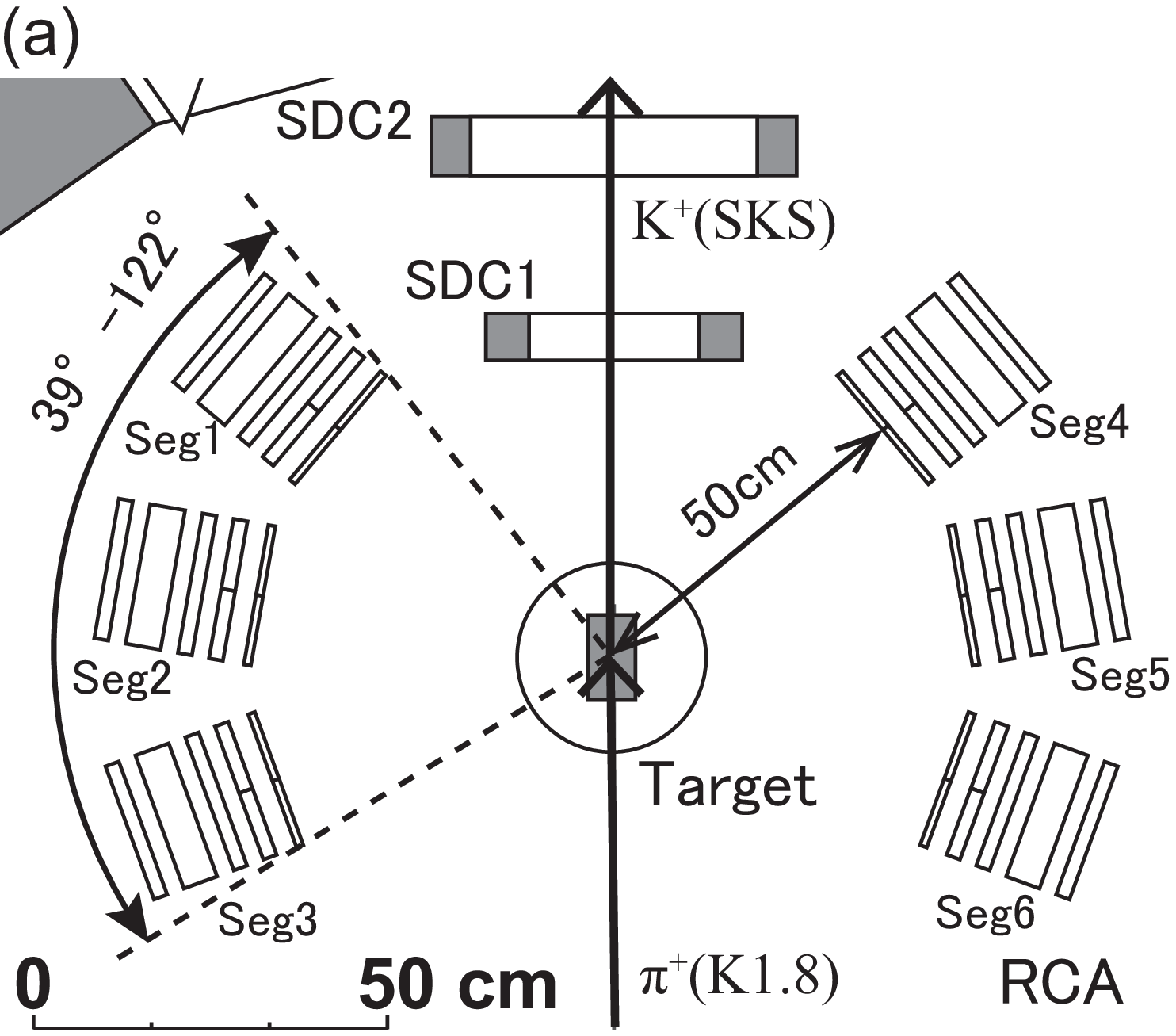} 
  \end{center}
 \end{minipage}
\hspace*{-0.2cm} 
 \begin{minipage}{0.5\hsize}
  \begin{center}
  \vspace*{-0.5cm}
   \includegraphics[width=80mm]{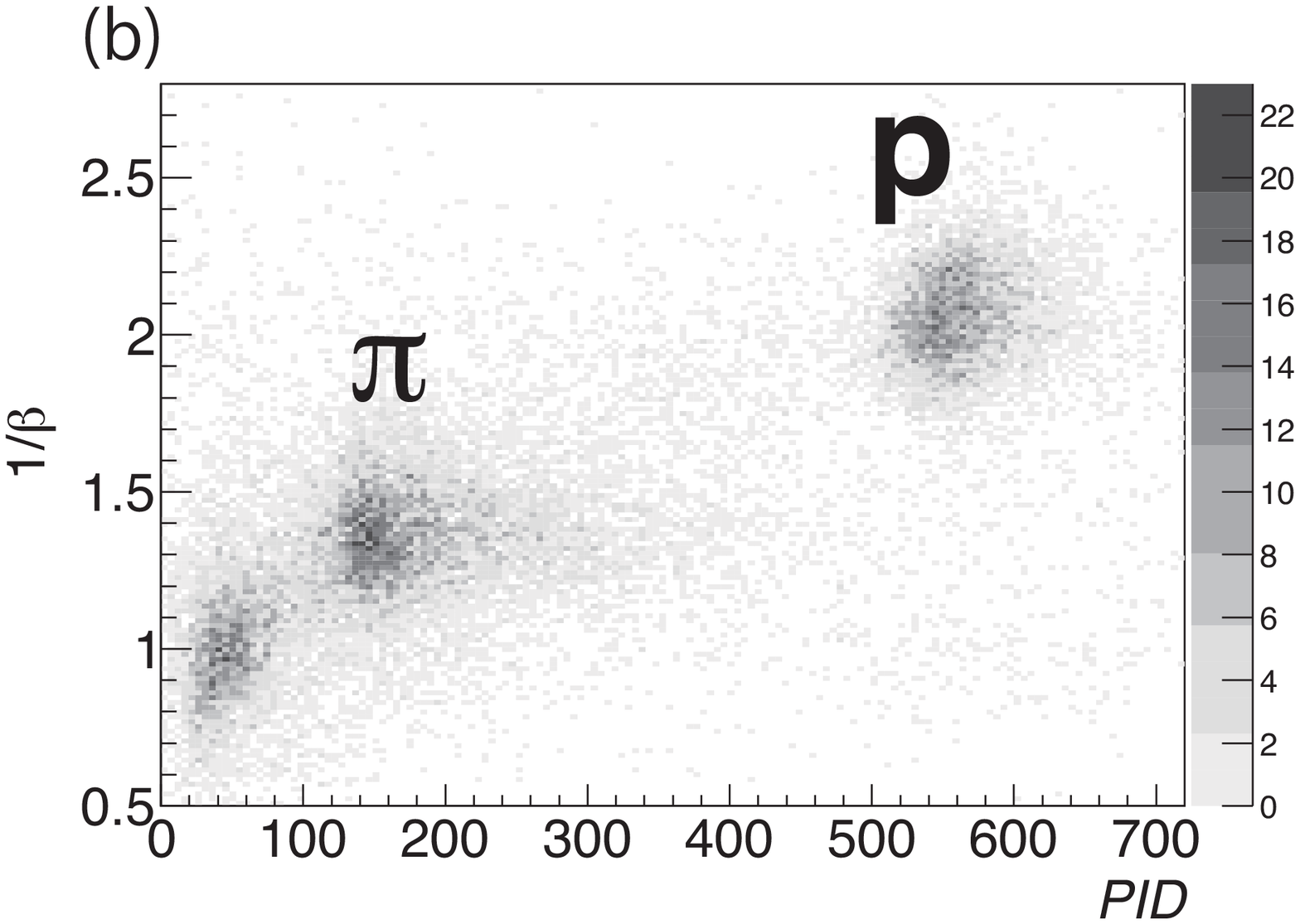}  
  \end{center} 
 \vspace*{-0.7cm} 
  \end{minipage}
 \caption{(a) Schematic view of the range counter system. It was
  composed of six range counter arrays; three in the left (Seg1 to Seg3)
  and three in the right (Seg4 to Seg6) of the beam axis. 
  SDC1 and 2 were the tracking drift chamber at the entrance of the SKS. 
  (b) A scatter plot between the $PID$ function and 1/$\beta$.
  Protons are clearly separated from pions. 
\label{fig:RCA}}
\end{figure}

In order to suppress a large backgrounds coming from
quasi-free productions of hyperons ($\Lambda$ and $\Sigma$'s) and 
hyperon resonances ($\Lambda(1405)$ and $\Sigma(1385)$'s),
a range counter system was installed surrounding the liquid
deuterium target in the laboratory angles between 39$^\circ$
and 122$^\circ$ both in the left and right sides from the beam
axis as shown in Fig.~\ref{fig:RCA}~(a). We had three range counter arrays (RCA's) in each side and  
the assignment of the segment number is also shown in Fig.~\ref{fig:RCA}~(a).

%
%
%
%

Each range counter array had five layers of plastic scintillation
counters; the thickness of each scintillator was  1~cm, 2~cm, 2~cm,
5~cm, and 2~cm, respectively,  with a height of 100~cm.
The width of each layer was 20~cm. The first two layers were
segmented into two slabs; each slab had 10-cm width.
Therefore, we had seven (2+2+1+1+1) scintillation counters in one range 
counter array. Every scintillation counter was read out
from both sides (up and down) by photo-multiplier
tubes (PMT).

%
From each PMT, both hit timing and pulse height information
were obtained. The discriminator threshold for the timing
information was set at less than the one tenth level of the minimum
ionizing particles.
The timing information from the first layer was used for the
on-line trigger and the time-of-flight analysis in off-line.
The distance from the liquid target center to the
first layer was about 50~cm.
In the on-line trigger, the $(\pi,K)$ trigger in coincidence with
range counter hits was generated by requiring at least one hit
among 12~first-layer scintillators.

From a hit pattern of five layers, we can define the stopping layer, $i_{stop}$,
for each range counter array.
Then, we set up a particle identification parameter, $PID$, as,
\begin{equation}
PID = (dE_{i_{stop}}+dE_{(i_{stop}-1)})^\alpha - (dE_{i_{stop}})^\alpha,
\end{equation}
where $dE_i$ shows the energy deposit in the $i$-th layer of
the plastic scintillators.
The $PID$ is a function of particle mass when the parameter
$\alpha$($\sim$1.75) is properly adjusted.

The time-of-flight (TOF) of each particle was obtained with the hit timing
in the first layer. 
The flight path length was measured from the 
vertex position of the $(\pi^+, K^+)$ reaction to the hit position on the first layer. 
In this analysis, the horizontal hit position was assumed to be the center 
of the scintillators and the vertical hit position was obtained with the 
time difference between the up and down PMT's. 
Then, the velocity of the particle ($\beta$) was obtained as 
$\beta$~=~(path~length)/(TOF$\cdot c$), which was adjusted by using the $\pi^+$ of $\beta\sim1$.

Thus, we used the $PID$ function and the velocity ($\beta$) for
the particle identification between proton and pion.
Fig.~\ref{fig:RCA}~(b) shows an example of a scatter
plot between $PID$ and 1/$\beta$ in the case of $i_{stop}=4$
in one of the RCA. 
In this analysis, the proton was selected as the gate of $\pm 3\sigma$ in PID and 
$\pm 2\sigma$ in 1/$\beta$ for each stopping layer.
Protons are clearly separated from pions. 
The energy of proton was determined from the velocity.

%
%
%
%

From a study of the hydrogen target data, we could identify the proton 
from the $\Sigma^+$ decay ($\Sigma^+ \to p\pi^0$) with a detection efficiency of 65 $\%$ 
for protons hitting the first layer of RCA.
The detection efficiency was limited due to a leakage
of particles at the side edges of RCA.


\section{3. Inclusive and Coincidence Analyses}

\begin{figure}
\begin{center}
\vspace{-0.8cm}
\hspace{-0.3cm}
\includegraphics[height=11cm]{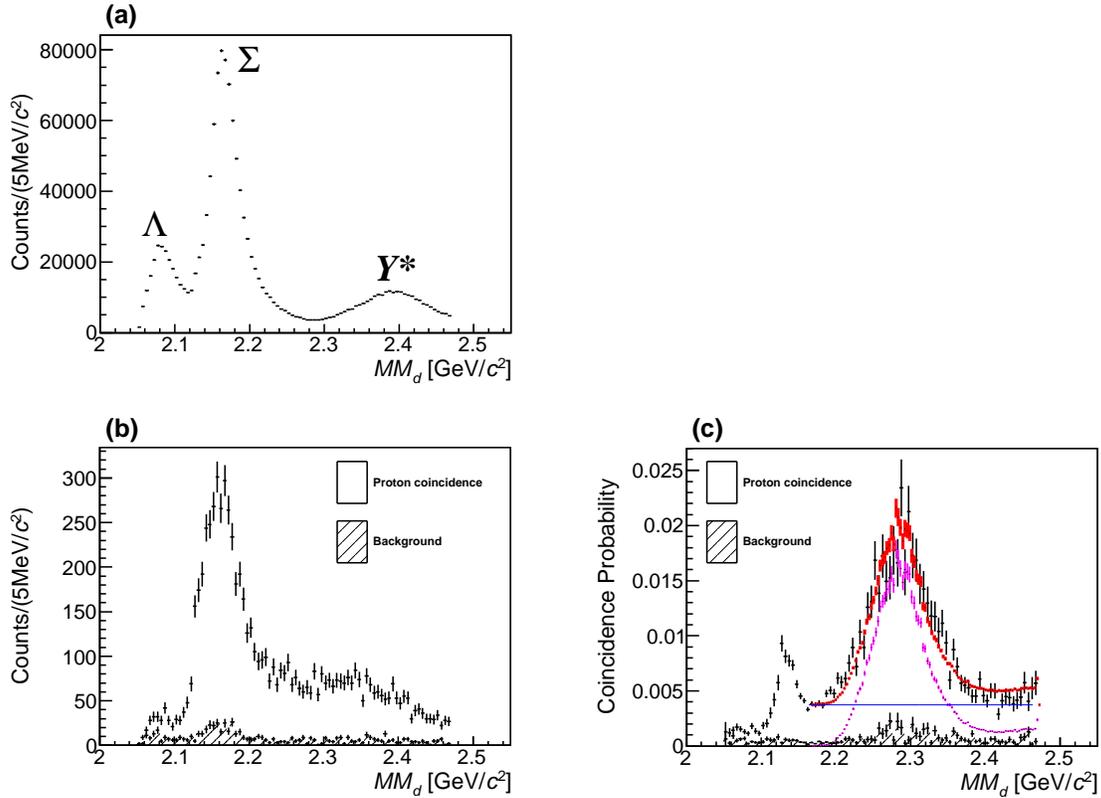}
\vspace{-1.1cm}
\caption{(a) Inclusive missing-mass spectrum of the $d(\pi^+, K^+)$ reaction
at 1.69~GeV/$c$ in the laboratory scattering angles from 2$^\circ$
to 14$^\circ$.
(b) Missing-mass spectrum of the $d(\pi^+, K^+)$ reaction
with one proton in the middle of the RCA in each side (Seg2, 5).
(c) The coincidence probability of a proton obtained by dividing the coincidence
spectrum (b) with the inclusive spectrum (a). 
Hatched spectra show the background contamination from the miss-identification of $\pi^{\pm}$ in RCA.
See the text in page \pageref{page_two_proton} for the detail of colored spectra in (c).
 \label{fig:Count_all}}
\end{center}
\end{figure}

Fig.~\ref{fig:Count_all}~(a) shows the inclusive missing-mass (MM$_d$)
spectrum for the $\pi^+d\rightarrow K^+X$ reaction at 1.69~GeV/$c$
in the scattering angles between 2$^\circ$ and 14$^\circ$ in the laboratory
frame, which is a spectrum without acceptance correction.~
The overall missing-mass resolution was estimated from the
missing-mass spectra of $\pi^+p\rightarrow K^+\Sigma ^+$
reactions at 1.58~GeV/$c$ and 1.69~GeV/$c$, and it was 2.8~$\pm$~0.1~MeV/$c^2$~(FWHM).
The details of the inclusive analyses were reported in Ref.~\cite{E27inclusive}.

As shown in Fig.~\ref{fig:Count_all}~(a), the missing-mass spectrum is composed of three major
components:
I)~the quasi-free $\pi^+n\rightarrow K^+\Lambda$ contribution (QF$\Lambda$),
I\hspace{-.1em}I)~the  quasi-free
$\pi^+p\rightarrow K^+\Sigma^+$ and 
$\pi^+n\rightarrow K^+\Sigma^0$ contributions (QF$\Sigma$),
and
I\hspace{-.1em}I\hspace{-.1em}I)~a mixture of the quasi-free
$\pi^+N\rightarrow K^+\Sigma(1385)$, 
$\pi^+n\rightarrow K^+\Lambda(1405)$ (QF$Y^*$), and
$\pi^+N\rightarrow K^+(\Lambda/\Sigma)\pi$ (QF$Y\pi$).
In between the QF$\Lambda$ and QF$\Sigma$, we have a small peculiar
structure corresponding to a threshold cusp for the 
$\Sigma N\rightarrow \Lambda N$ conversion process.

%
%
%
 
%


%
%
%
%
Next, we request coincidence of one proton. 
According to our detector simulation, a proton emitted from QF$\Lambda$, QF$\Sigma$, QF$Y^*$, and QF$Y\pi$
processes rarely hits the RCA; only small fraction in the very forward
segments (Seg1,~4). 
The spectator proton in a deuteron rarely exceeds
the analysis threshold momentum of 250~MeV/$c$. Therefore,
we can expect a good suppression of the quasi-free processes
in one-proton coincidence spectrum. 

%
%
%
%

Fig.~\ref{fig:Count_all} (b) shows a coincidence spectrum
with one proton in the middle segments of the RCA in each side (Seg2,~5), 
where these segments have an almost flat and wide acceptance in missing mass. 
Note that there would be no quasi-free contributions in this
spectrum according to the simulation.
Possible non-quasifree contributions are
the threshold cusp emitting through a strong conversion of $\Sigma^+n\rightarrow\Lambda p$,
the $K^-pp$ signal emitting through $K^-pp\rightarrow\Lambda(\Sigma^0)p$, 
and quasi-free hyperons and hyperon resonance productions
followed by 
conversions such as $\Sigma N\rightarrow\Lambda N$. 

In Fig.~\ref{fig:Count_all}~(c), we present a ratio histogram
between the one-proton coincidence spectrum (Fig.~\ref{fig:Count_all}~(b)) and the
inclusive one (Fig.~\ref{fig:Count_all}~(a)). 
This is a spectrum without acceptance correction for the RCA.
The vertical axis shows
the proton coincidence probability as a function of the missing mass.
The background contamination from the miss-identification of $\pi^{+/-}$ in RCA, 
which is estimated by the side-band events in $PID$, is shown with hatched spectra in Figs.~\ref{fig:Count_all}~(b) and (c). 
The contamination fraction of this background is about 7$\%$ around MM$_{d} \sim $2.27~GeV/$c^2$. 

We notice there are two prominent structures; one at the
threshold cusp position~(2.13~GeV/$c^2$) and the other broad 
bump at around 2.27~GeV/$c^2$, which can be a signal
of the ``$K^-pp$"-like structure.
In the QF$\Sigma$ and QF$Y^*$ region, the proton coincidence
probability is smaller than the two prominent structures and
stays rather constant. 

%
%
%

\begin{figure}
\begin{center}
 \vspace{-0.5cm}
 \includegraphics[height=6cm]{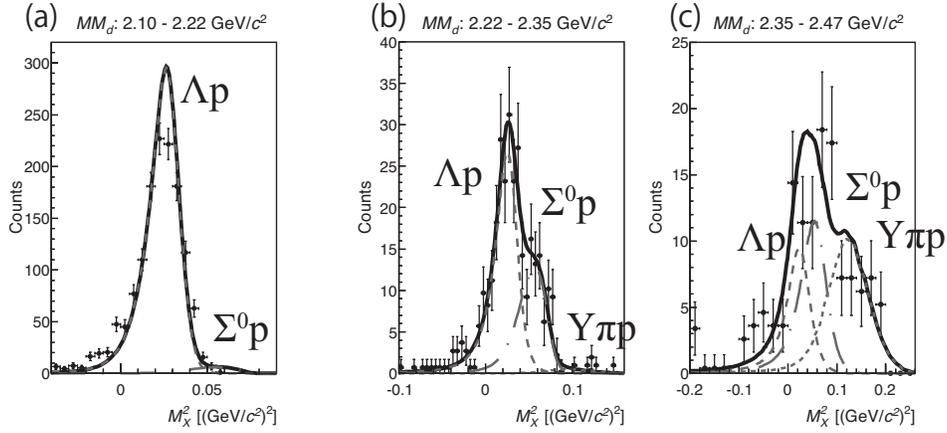}
 \vspace{-0.4cm}
  \caption{Missing-mass square spectra of $X$ obtained in the two-protons coincidence
  events in the reaction of the $d(\pi^+, K^+pp)X$.
    Each spectrum shows the mass square of $X$ for different $MM_{d}$ region; 
    the left (a) shows the QF$\Sigma$ 
    region~($MM_{d}<2.22~$GeV/$c^{2}$), the center (b) shows ``$K^-pp$"-like structure   
    region~($2.22<MM_{d}<2.35~$GeV/$c^{2}$)  and the right (c) shows the QF$Y^{*}$ 
    region~($MM_{d}>2.35~$GeV/$c^{2}$). 
    The spectra were fitted with three components of $\Lambda p$~(dashed~line), $\Sigma^0 p$~(dot-dashed~line) and 
    $Y \pi p$~(dotted~line) decay modes.
   \label{fig:Xmass}}
\end{center}
\end{figure}

At this stage, the acceptance of our range counter system is
not taken into account.
The acceptance correction needs information of decay modes of the
``$K^-pp$"-like structure. This study was carried out by requiring
coincidence of two protons in the RCA's.
In such a condition, we can measure the missing-mass of $X$ in the $d(\pi^+, K^+pp)X$
process by detecting two protons in the decay of the $ppX$ system, of which mass is $MM_d$, 
in three categories;
a)~$\Lambda p,~\Lambda\rightarrow p\pi^-$, 
b)~$\Sigma^0p,~\Sigma^0\rightarrow\Lambda\gamma\rightarrow p\pi^-\gamma$, 
and 
c)~$Y\pi$N$\rightarrow pp\pi\pi$.
The first two modes, a) and b), are non-mesonic and the $X$ is one pion (and $\gamma$).
The last one, c), is mesonic and the $X$ is two pions. 
Therefore, the missing-mass spectrum of $M_X$ 
should show different distributions for each decay mode.
Fig.~\ref{fig:Xmass} shows such missing-mass square spectra of $M_X$.
Three distributions estimated for each decay mode are shown in
the figure by fitting the height of each template distribution. 
These templates were made from the simulation, which assumes the reaction of 
$\pi^+d \to K^+W, W \to pY (p(Y\pi))$ with uniform productions and decays in the center of mass system.

%
%
%

\begin{figure}
\begin{center}
  \vspace{-0.69cm}
  \includegraphics[height=5.5cm]{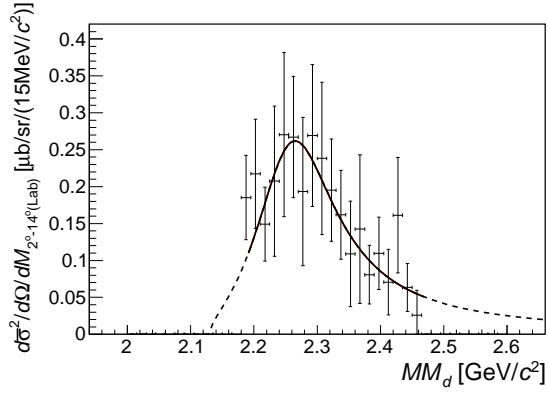}
  \vspace{-0.42cm}
  \caption{Missing-mass spectrum of the $d(\pi^+, K^+)$ reaction
  for two-protons coincidence and the $\Sigma^0 p$ decay branch events. 
  The mass acceptance of the RCA's are
  corrected. 
  The spectrum was fitted with a relativistic Breit-Wigner function 
  (see the text for the detail).
  We found the mass 2275~$^{+17}_{-18}$~(stat.)~$^{+21}_{-30}$~(syst.)~MeV$/c^2$    
and the width 162~$^{+87}_{-45}$~(stat.)~$^{+66}_{-78}$~(syst.)~MeV.\label{fig:TWOPcoincidence}}
\end{center}
\end{figure}

According to these fitting results, we can correct the acceptance of the RCA for each decay mode. 
It is almost flat in the missing mass except near the
threshold for each decay mode.
Fig.~\ref{fig:TWOPcoincidence} shows a missing-mass distribution
for two-protons coincidence events of the $\Sigma^0$p final state~b) with the acceptance correction.
The spectrum was fitted with a relativistic Breit-Wigner function as, 
\begin{equation}
f(MM_d) = \frac{(2/\pi)MM_dm_0\Gamma(q)}{(m_0^2-MM_d^2)^2+(m_0\Gamma(q))^2}. \label{eqn:relaBW}
\end{equation}
The mass-dependent width was $\Gamma(q) =\Gamma_0(q/q_0)$, in which $q~(q_0)$ is the momentum 
of the $\Sigma^0$ and proton in the $\Sigma^0 p$ rest frame at mass $MM_{d}$~($m_0$). 
The obtained mass and width are 2275~$^{+17}_{-18}$~(stat.)~$^{+21}_{-30}$~(syst.)~MeV$/c^2$  
and 162~$^{+87}_{-45}$~(stat.)~$^{+66}_{-78}$~(syst.)~MeV, respectively.
It corresponds to the biding energy of the $K^-pp$ system to be 
95~$^{+18}_{-17}$~(stat.)~$^{+30}_{-21}$~(syst.)~MeV and the production cross section of 
the ``$K^-pp$"-like structure decaying to $\Sigma^0p$ of 
$d\sigma /d\Omega_{K^-pp \to \Sigma^0p}$ = 4.4~$\pm$~0.4~(stat.)~$^{+0.8}_{-1.6}$~(syst.)~$\mu b/sr$. 
The systematic errors of these values were estimated taking into account uncertainties 
in the fitting ranges, the binning of the missing-mass spectrum, 
the detection efficiency of two protons in RCA and 
the Breit-Wigner shape by changing the Lorentzian function folded with the missing-mass resolution.     
The differential cross section of ``$K^-pp$"-like structure of the $\Lambda p$ decay mode~a) 
was also estimated from the fitting assuming the same distribution of $MM_d$. 
Thus, a branching fraction of the ``$K^-pp$"-like structure was obtained to be   
$\Gamma_{\Lambda p}/\Gamma_{\Sigma^0 p}$ = 0.73~$^{+0.13}_{-0.11}$~(stat.)~$^{+0.43}_{-0.21}$~(syst.).
This ratio was discussed from a theoretical point in Ref.~\cite{Sekihara}.

Next, we try to understand the ratio histogram (Fig.~{\ref{fig:Count_all}}~(c)) with the obtained
$K^-pp$ mass distribution of $f(MM_d)$.~\label{page:twoproton}
By using the mass distribution for the ``$K^-pp$"-like structure 
and the double differential cross section
of the inclusive $(\pi^+, K^+)$ process $\frac{d^2\sigma}{d\Omega dMM_d}(MM_d)_{Inclusive}$,
we can obtain the ratio histogram as shown in Fig.~{\ref{fig:Count_all}}~(c) as a plot colored in pink,
which is calculated as,
\begin{equation}
R_p(MM_d) = \frac{C\times f(MM_{d})\times \eta_{1p}(MM_d)}{(\frac{d^2\sigma}{d\Omega dMM_d}(MM_d))_{Inclusvie}},
\end{equation}
where $C$ is the normalization constant, 
and $\eta_{1p}(MM_d)$ is the detection efficiency of a proton in the middle segments of the RCA (Seg2, 5). 
A blue line in Fig.~{\ref{fig:Count_all}}~(c) is an assumed  flat component representing the 
conversion processes and the background contamination 
from the miss-identification of $\pi^{\pm}$ in RCA. 
Red points with error bars in Fig.~{\ref{fig:Count_all}}~(c) are the sum of the pink points and blue line. 
The normalization constant $C$ and the amplitude of the flat component (blue line) 
were adjusted to minimize the differences between the black and red points. 
Thus, the obtained one proton coincidence probability spectrum of the broad enhanced region could 
be reproduced by the ``$K^-pp$" and flat background. 
\label{page_two_proton}

%
%
%

What is the nature of the ``$K^-pp$”-like structure? 
It should have strangeness $-1$ and baryon number $B = 2$ from the observed reaction mode, 
so that the hyper charge $Y = 1$.
As for the spin of the $K^-pp$ system, a $K^-$ is theoretically assumed 
to couple with a spin-singlet ($S = 0$) 
p-p pair in $S$-wave~($L = 0$). 
So that the $J^{P} = 0^-$, presumably.
Alternative view of the system as a $\Lambda^*p$ 
bound state~\cite{UCHINO} also predicts the bound state spin to be 0. 
There is also a  theoretical prediction of a $(Y, I, J^P ) = (1, 3/2, 2^+)$ 
dibaryon as $\pi\Lambda N \mbox{--} \pi\Sigma N$ bound state~\cite{GG}.



\section{4. Summary}
We have observed a ``$K^-pp$"-like structure in the $d(\pi^+,K^+)$
reaction at 1.69~GeV/$c$ with coincidence of high-momentum ($>$250~MeV/$c$) proton(s)
in large emission angles ($39^\circ<\theta_{lab.}<122^\circ$).
A broad enhancement in the proton(s) coincidence spectra are observed
around the missing-mass of 2.27~GeV/$c^2$, which corresponds
to the binding energy of the $K^-pp$ system of 
95~$^{+18}_{-17}$~(stat.)~$^{+30}_{-21}$~(syst.)~MeV and
the width of 162~$^{+87}_{-45}$~(stat.)~$^{+66}_{-78}$~(syst.)~MeV. 
The branching fraction between the $\Lambda p$ and $\Sigma^0 p$ decay modes of the ``$K^-pp$"-like structure was
measured to be 
$\Gamma_{\Lambda p}/\Gamma_{\Sigma^0 p}$ = 0.73~$^{+0.13}_{-0.11}$~(stat.)~$^{+0.43}_{-0.21}$~(syst.), 
for the first time. 

\ack
We would like to thank the Hadron beam channel group, 
accelerator group and cryogenics section in J-PARC for their great efforts on
stable machine operation and beam quality improvements.
The authors thank the support of NII for SINET4.
This work was supported by the Grant-In-Aid for Scientific
Research on Priority Area No. 449 (No. 17070005), 
the Grant-In-Aid for Scientific Research on Innovative Area
No. 2104 (No. 22105506), from the Ministry of Education,
Culture, Sports, Science and Technology (MEXT) Japan, and 
Basic Research (Young Researcher) No. 2010-0004752 from 
National Research Foundation in Korea.
We thank supports from National Research Foundation, WCU program of the
Ministry of Education, Science and Technology (Korea), Center for Korean
J-PARC Users.

\end{sloppypar}
\end{document}